\documentclass[11pt,letterpaper]{article}
\pdfoutput=1

\usepackage{jheppub}
\usepackage{color}
\usepackage{graphicx}
\usepackage{wrapfig,enumerate,slashed}
\usepackage[utf8]{inputenc}
\usepackage{adjustbox}

\usepackage{footmisc}
\usepackage{amsmath}
\usepackage{wasysym} 
\usepackage{graphicx}
\usepackage{color}
\usepackage{comment}
\usepackage{hyperref}
\usepackage{subcaption}
\usepackage{slashed}
\usepackage{booktabs}

\newcommand{\be}{\begin{eqnarray}}
\newcommand{\ee}{\end{eqnarray}}

\newcommand{\bee}{\begin{eqnarray}}
\newcommand{\eee}{\end{eqnarray}}
\newcommand{\beeq}{\begin{equation}}
\newcommand{\eeeq}{\end{equation}}

\begin{document}

\title{Adversarially-trained autoencoders for robust unsupervised new physics searches}

\author[a,b]{Andrew Blance,}
\author[a]{Michael Spannowsky,}
\author[a]{and Philip Waite}

\affiliation[a]{Institute for Particle Physics Phenomenology, Department of Physics, Durham University, Durham, DH1 3LE, UK}

\affiliation[b]{Institute for Data Science, Durham University, Durham, DH1 3LE, UK}

\emailAdd{andrew.t.blance@durham.ac.uk}
\emailAdd{michael.spannowsky@durham.ac.uk}
\emailAdd{p.a.waite@durham.ac.uk}

\abstract{Machine learning techniques in particle physics are most powerful when they are trained directly on data, to avoid sensitivity to theoretical uncertainties or an underlying bias on the expected signal. To be able to train on data in searches for new physics, anomaly detection methods are imperative, which can be realised by an autoencoder acting as an unsupervised classifier. The last source of uncertainties affecting the classifier are then experimental uncertainties in the reconstruction of the final-state objects. To mitigate their effect on the classifier and to allow for a realistic assessment of the method, we propose to combine the autoencoder with an adversarial neural network to remove its sensitivity to the smearing of the final-state objects. We quantify its effect and show that one can achieve a robust anomaly detection in resonance-induced $t\bar{t}$ final states.}

\preprint{IPPP/19/41}

\maketitle


\section{Introduction}
\label{sec:intro}
In recent years machine learning algorithms, and in particular neural networks, have become increasingly popular in analysing large quantities of data. In the context of particle physics two main applications are prevalent: the classification of data according to different hypotheses \cite{Nachman:2014kla, Komiske:2016rsd,Barnard:2016qma, Dery:2017fap, Butter:2017cot,Cohen:2017exh,Chang:2017kvc,Pearkes:2017hku,Louppe:2017ipp,Kasieczka:2017nvn, deOliveira:2017pjk,Luo:2017ncs,Datta:2017lxt,Larkoski:2017jix,Shimmin:2017mfk,Metodiev:2017vrx,Roxlo:2018adx,Brehmer:2018kdj,Brehmer:2018eca,Collins:2018epr,Duarte:2018ite,Fraser:2018ieu,Komiske:2018oaa,Macaluso:2018tck,Andreassen:2018apy,deCastro:2018mgh,DAgnolo:2018cun,Brehmer:2018hga,Monk:2018zsb,Moore:2018lsr,DeSimone:2018efk,Bollweg:2019skg,Cerri:2018anq} and the regression of data to interpolate and extrapolate object-relevant properties \cite{ATL-PHYS-PUB-2018-013,Aaboud:2017all,Khachatryan:2014gga,Khachatryan:2015hwa,Khachatryan:2015iwa}. 

Using multi-variate analysis (MVA) techniques to classify events into signal and background classes based on their radiation profiles should improve the LHC's experiments' sensitivity in searches for new physics. Machine learning algorithms are able to analyse multiple observables or inputs simultaneously to find a region in this multi-dimensional parameter space that shows a relative enhancement of signal over background events. To find this region in a supervised-learning approach, pseudo-data for signal and background need to be generated using event generators, e.g.~\textsc{Sherpa} \cite{Gleisberg:2008ta}, \textsc{Herwig} \cite{Bellm:2015jjp} or \textsc{Pythia} \cite{Sjostrand:2014zea}, and the respective training samples are made known to the algorithm whether they contain signal or background respectively. However, as the Monte Carlo event samples are plagued by theoretical uncertainties, the classification algorithm will be subjected to the same uncertainties. This issue is amplified by the fact that the MVA method will usually favour highly-exclusive phase space regions which are poorly understood perturbatively \cite{Englert:2017aqb,Englert:2019xhk}, and often observables that are not even IR-safe are found in experimental measurements to be most discriminative, e.g.~the number of charged tracks \cite{Schaetzel:2013vka,Aad:2014gea}, thus further questioning the reliability of theoretically predicted classification efficiencies. Adversarial neural networks have been proposed to desensitise classification methods against theoretical \cite{Englert:2018cfo} and systematic uncertainties \cite{Louppe:2016ylz} or against certain observables \cite{Heimel:2018mkt}. 
 
One promising approach to overcome deficits from training on pseudo-data is to train on actual data directly.\footnote{If machine learning techniques can be trained on data directly they become independent of theoretical uncertainties. In such circumstances they can outperform theory-based reconstruction approaches, like the matrix element method \cite{Kondo:1988yd,Abazov:2004cs, Abulencia:2005pe,Artoisenet:2010cn,Martini:2015fsa}, which was recently extended to fully exclusive final states \cite{Soper:2011cr, Soper:2012pb,Soper:2014rya,Englert:2015dlp,FerreiradeLima:2017iwx,Prestel:2019neg}.} While so-called data-driven methods are not subjected to theoretical uncertainties, one has to make sure that signal and background are sufficiently pure to train the algorithm on well-separated event samples. Most of the time, and in particular in searches for new physics, this is a highly challenging task. Rare processes, e.g.~the production of di-Higgs final states, or completely unknown processes, e.g.~the production of a gluino, are of utmost interest to search for at the LHC. However, obtaining a data-driven training sample for such processes is impossible, thus, limiting the applicability for data-driven methods. One way around this bottleneck is not to train on signal at all, but to identify the kinematic features of background samples and to design a method that flags up events that do not possess the same features, thereby classifying such an event as signal. The remaining residual experimental problem that remains for such an approach are the experimental and systematic uncertainties in the measurements of the inputs of a data-driven anomaly detection method.

Autoencoders \cite{bangalore, Kingma:2013hel} have been proposed for denoising \cite{Vincent:2008:ECR:1390156.1390294}, generative models \cite{Otten:2019hhl} and in particular for anomaly detection \cite{Farina:2018fyg, Hajer:2018kqm,Heimel:2018mkt,Roy:2019jae}. They use an information bottleneck to map an input to a latent-compressed representation and then decode this representation back. The loss function measures the squared difference between input and decoded output. By minimising the loss function, the autoencoder learns intrinsic features of the training samples that survive the information bottleneck. After training the autoencoder on background samples, it is expected that applying the autoencoder to signal samples will result in a modified value for the loss function, as some kinematic features differ between signal and background. The depth of the networks and the width of the bottleneck are hyperparameters of the network that can be optimised for the problem at hand. Using autoencoders for anomaly detection, we will show that adversarially-trained neural networks can take systematic uncertainties into account and desensitise the classification performance in data-driven searches for new physics. To achieve this, we adversarially train an autoencoder on Monte-Carlo-generated pseudo-data which has been systematically smeared in order for it to learn to reconstruct the events without using any information about the smearing.

We apply this framework to resonance searches, i.e.~a heavy colour-singlet scalar, colour-octet scalar and colour-singlet vector, that are well-motivated by many new physics models. This selection allows one to study the impact of the spin and colour quantum numbers of the resonances on the classification efficiencies.\footnote{The quantum numbers of the decaying resonances are known to have a strong impact on the reconstruction efficiencies of boosted top quarks \cite{Joshi:2012pu}.} The resonances are assumed to subsequently decay into top quarks \cite{Chatrchyan:2012ku, Aaboud:2018mjh,Aaboud:2019roo,Aaboud:2017hnm}. Top quark samples are an ideal playground for anomaly detection, as they can be purified to a very high degree, i.e.~the confidence that one trains on a pure $t\bar{t}$ sample is very high, in particular when one top decays hadronically while the other decays leptonically. On the other hand, top final states are complex, consisting of many jets, leptons and missing transverse energy. Thus, uncertainties on reconstructed observables due to detector effects can be large.

The paper is structured as follows: in Sec.~\ref{sec:setup} we first discuss the analysis setup. To establish a baseline of what can be achieved by supervised learning, we show the performance of a neural network classifier and the effect of combining it with an adversarial neural network in Sec.~\ref{sec:supervised}. In Sec.~\ref{sec:unsupervised} we extend this approach to an unsupervised autoencoder for anomaly detection and consider its application to other new physics models and the effects of an impure training sample. We offer conclusions on our findings in Sec.~\ref{sec:conclusions}.

\section{Analysis setup and smearing procedure}
\label{sec:setup}

We use \textsc{MadGraph5\_{}aMC@NLO} \cite{Alwall:2014hca} to generate the events for the study, followed by \textsc{Pythia 8.2} for parton shower and hadronisation. The background events consist of $pp \rightarrow t\bar{t}$ at a centre-of-mass energy of $14$~TeV, with one top quark forced to decay leptonically and the other hadronically. The signal events are generated from a heavy $Z^\prime$ boson \cite{Altarelli:1989ff} via $pp \rightarrow Z^\prime \rightarrow t\bar{t}$, also with semileptonic decays of the top quarks. As a benchmark for this study, we select the $Z^\prime$ mass to be $2$~TeV with a width of $89.6$~GeV. A transverse momentum cut of $p_T > 500$~GeV is applied directly to the top quarks at generator level, for both signal and background events. 

Following the concept of reconstructing highly boosted top quarks with fat jets \cite{Plehn:2011tg,Plehn:2011sj}, the hadrons and non-isolated leptons from the event are initially clustered into jets using the Cambridge-Aachen algorithm \cite{Dokshitzer:1997in} with a radius of $R=1.0$. The constituents of the two hardest fat jets are then reclustered into jets using the $k_T$ algorithm with $R=0.2$, implemented in \textsc{FastJet} \cite{Cacciari:2011ma}. Jets are required to have $p_T > 30$~GeV and are $b$-tagged through their association to a $B$-meson. Isolated leptons are required to have $p_T > 10$~GeV. Events are selected which have a scalar-summed visible transverse momentum of $H_T>1$~TeV, and which have at least one $b$-jet inside one fat jet, at least one $b$-jet and two light jets inside the other fat jet, and at least one isolated lepton.

The observables that we consider for the analysis are the four-momenta of the two $b$-jets, two light jets and isolated lepton, as well as the missing energy ($\slashed{E}_T$) in the event (21 observables in total).
To represent possible systematic uncertainties that can arise in detectors from jet energy scales, we apply a smearing procedure to the jets and the missing energy in the events.
For the jets and leptons, we use a smearing based on Refs.~\cite{Buckley:2010ar, ATLAS:2015uwa} where the three-momenta of each object is smeared with a Gaussian. In our case, we take the extremities of this Gaussian so that the smearing is either applied upwards or downwards for all objects, with the relative width of the smearing envelope being larger for smaller $p_T$ values. Similarly, we apply a shift to the missing energy based on Ref.~\cite{Aad:2012re}, where the width of the shift is proportional to $\sqrt{H_T}$, and use the two extremities of the envelope. We fix the direction of the missing energy smearing to always be the same as that of the jets and leptons. For the purposes of this study, we increase the size of the smearing envelope by a further factor of three, to be conservative on the systematic uncertainties and highlight the ability of our setup to correct for it.

We apply the smearing to the background sample such that two extra datasets are created for smearing in the upwards and downwards directions, as well as the unsmeared central sample. No smearing is applied to the signal sample. The three background samples are each generated from statistically independent generator samples, and after all cuts we select 100,000 events from each of the four samples, with 20\% of these retained for testing.

\begin{figure}[!t]
\centering
   \begin{subfigure}{0.49\linewidth} \centering
     \includegraphics[width=\textwidth]{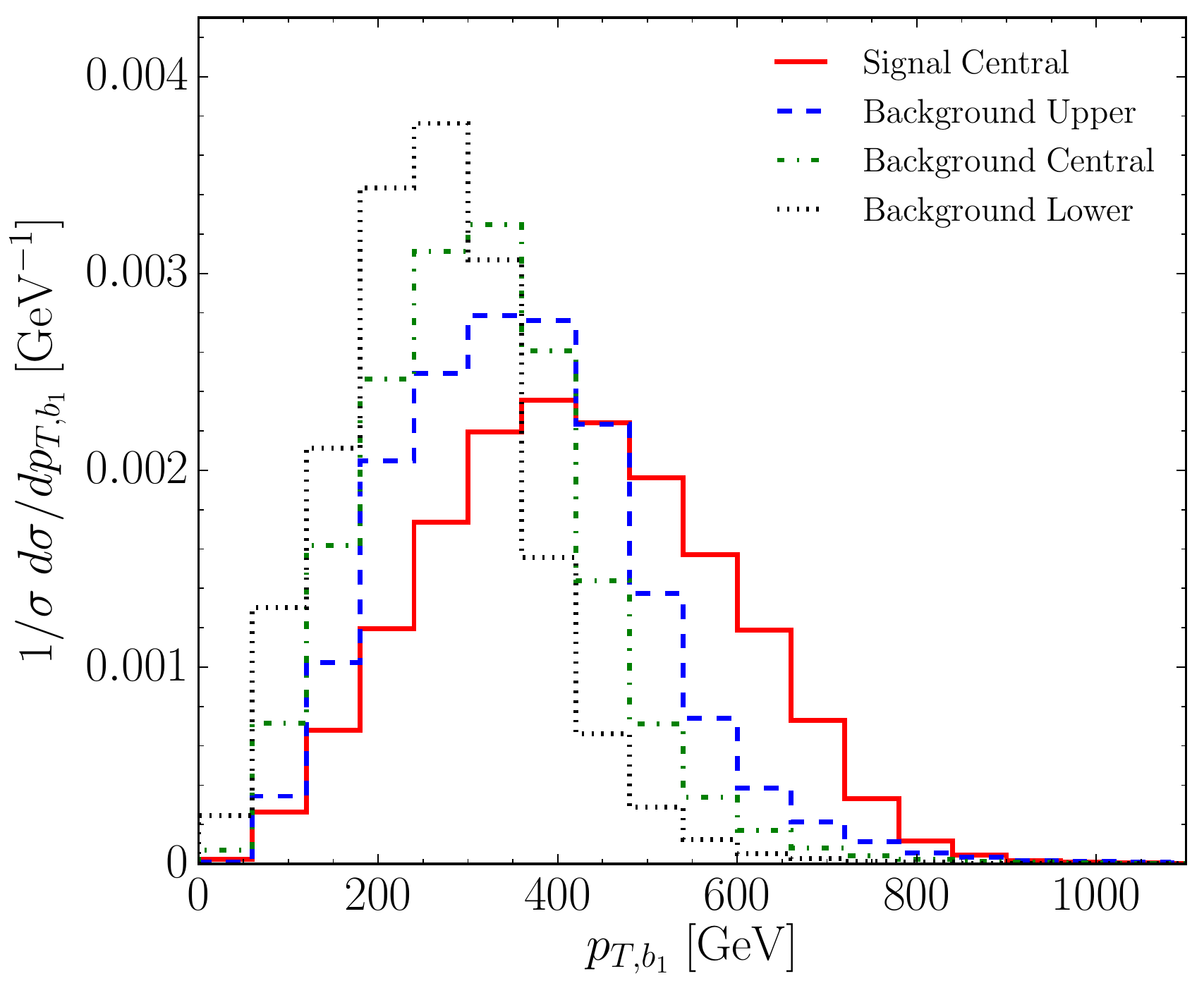}
     \caption{}
   \end{subfigure}
   \begin{subfigure}{0.49\linewidth} \centering
     \includegraphics[width=\textwidth]{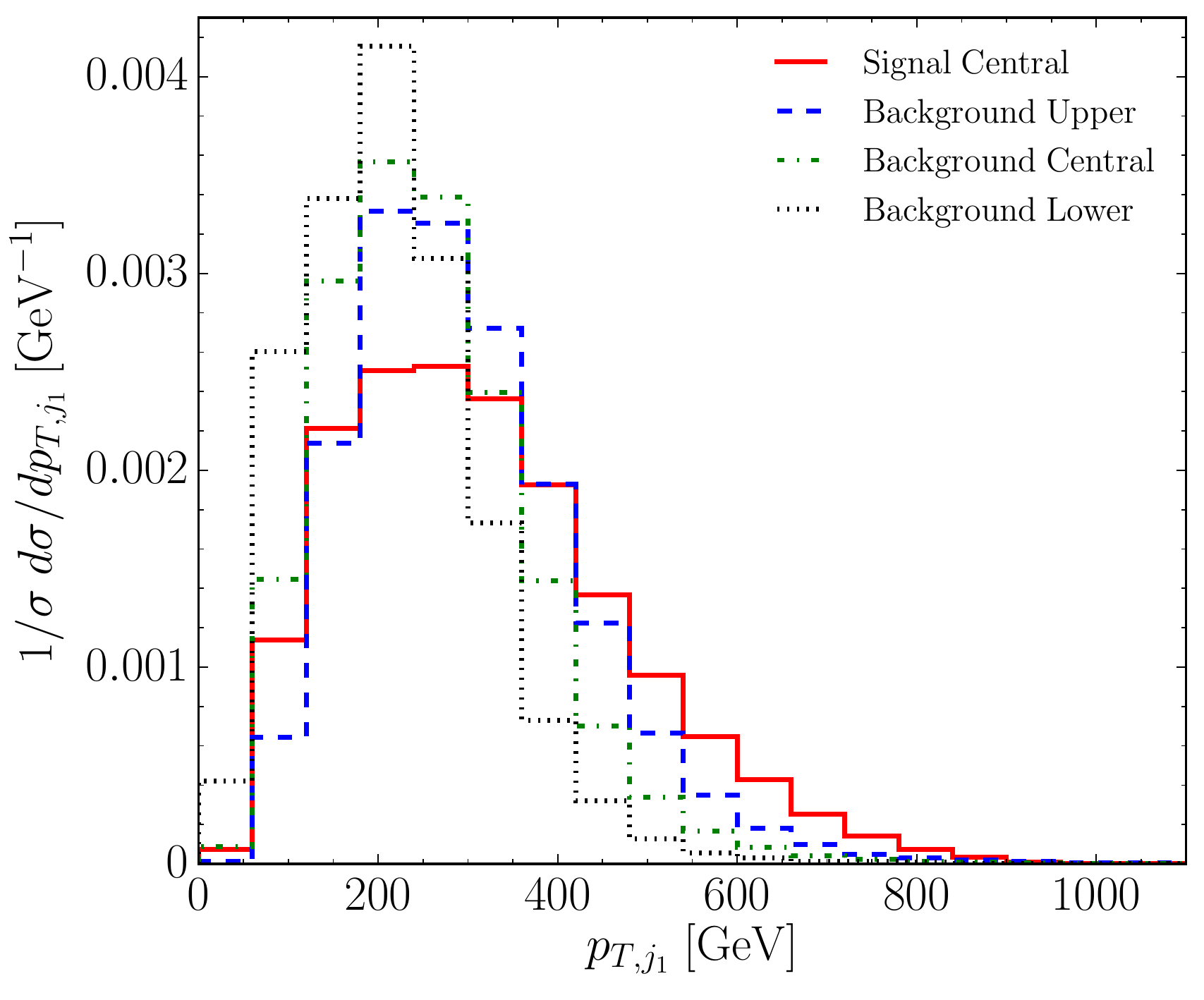}
     \caption{}
   \end{subfigure}
   \begin{subfigure}{0.49\linewidth} \centering
     \includegraphics[width=\textwidth]{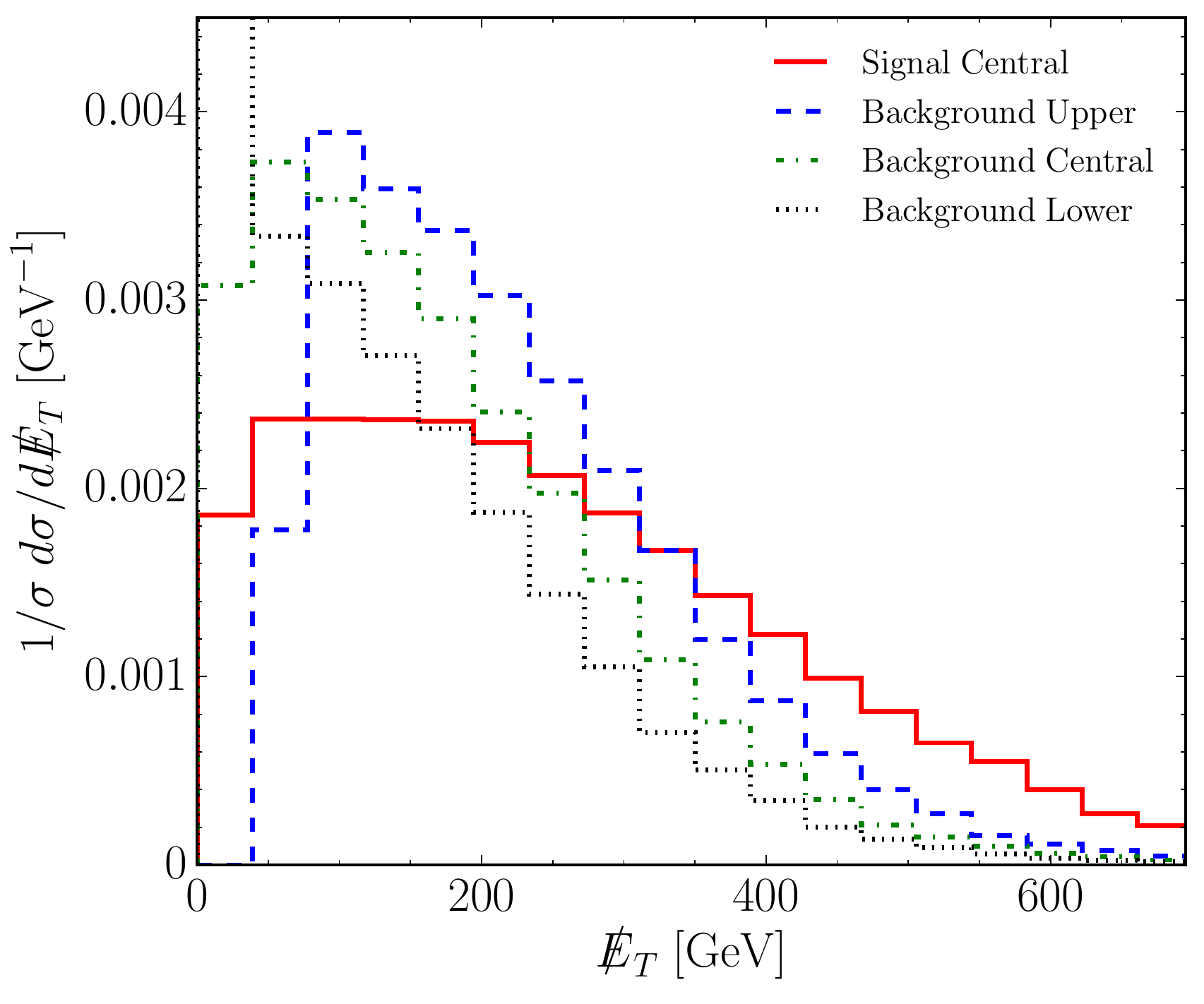}
     \caption{}
   \end{subfigure}
   \begin{subfigure}{0.49\linewidth} \centering
     \includegraphics[width=\textwidth]{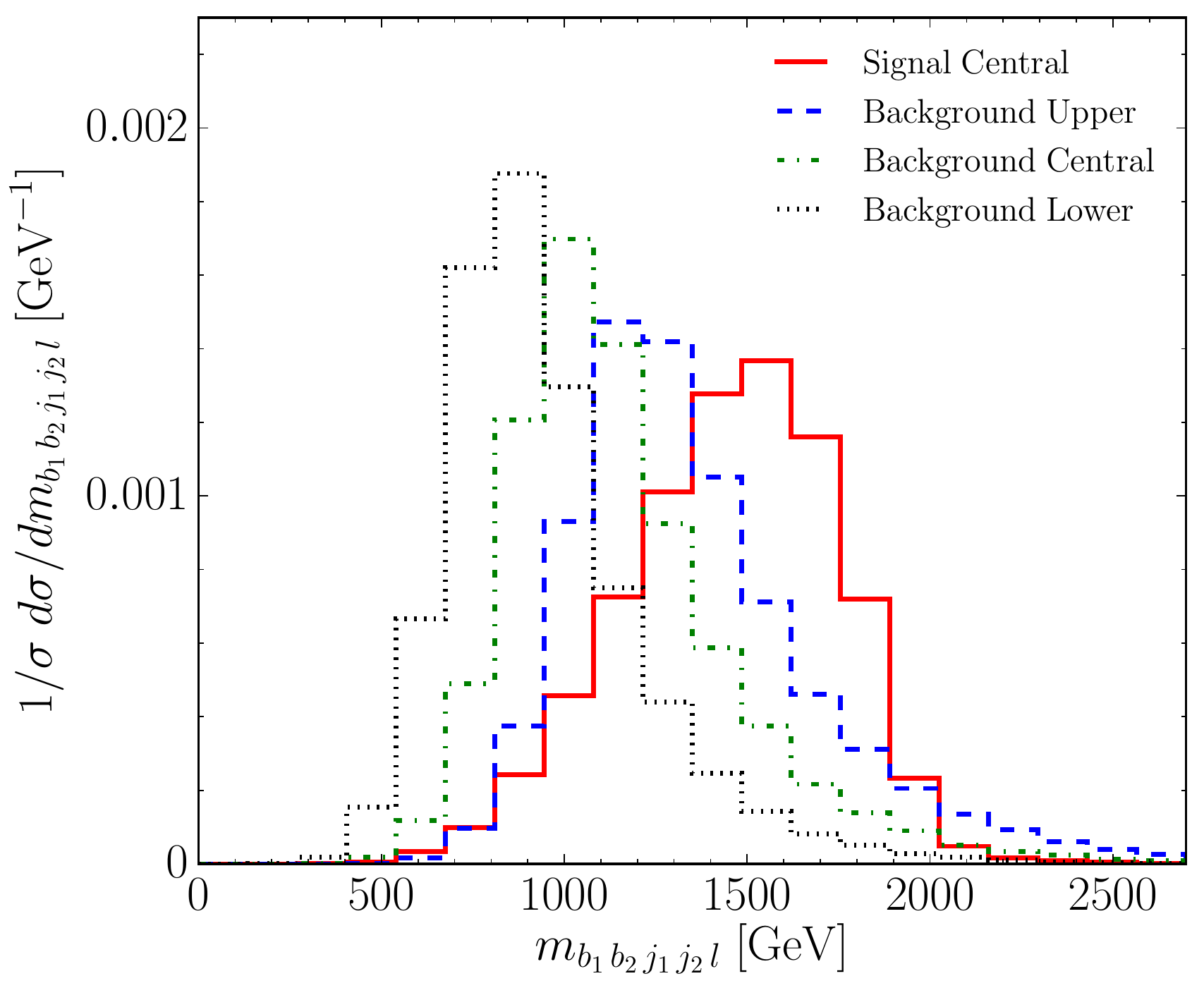}
     \caption{}
   \end{subfigure}
\caption{Effect of smearing on (a) the $p_T$ of the hardest $b$-jet, (b) the $p_T$ of the hardest light jet, (c) the missing energy and (d) the invariant mass of the jets and lepton, compared to the unsmeared background and the signal samples.} \label{fig:observables}
\end{figure}

In Fig.~\ref{fig:observables} we show the effect of smearing on the $p_T$ of the hardest $b$-jet and light jet, the missing energy in the event and the invariant mass of the jets and lepton, compared to the equivalent distributions for the signal events. Clearly the smearing of the background has the potential to make it either easier or harder for a classifier to discriminate between signal and background, depending on which direction the smearing shifts the background distribution.

\section{Decorrelated jet smearing with supervised adversarial classifier}
\label{sec:supervised}

To set a benchmark for the signal-to-background separation, we first train a simple neural network classifier to discriminate signal events from the complete set of background events (including all three samples). We expect this supervised-learning approach to perform better than the unsupervised approach which follows. 

\begin{figure}[!t]
\centering
   \begin{subfigure}{0.485\linewidth} \centering
     \includegraphics[width=\textwidth]{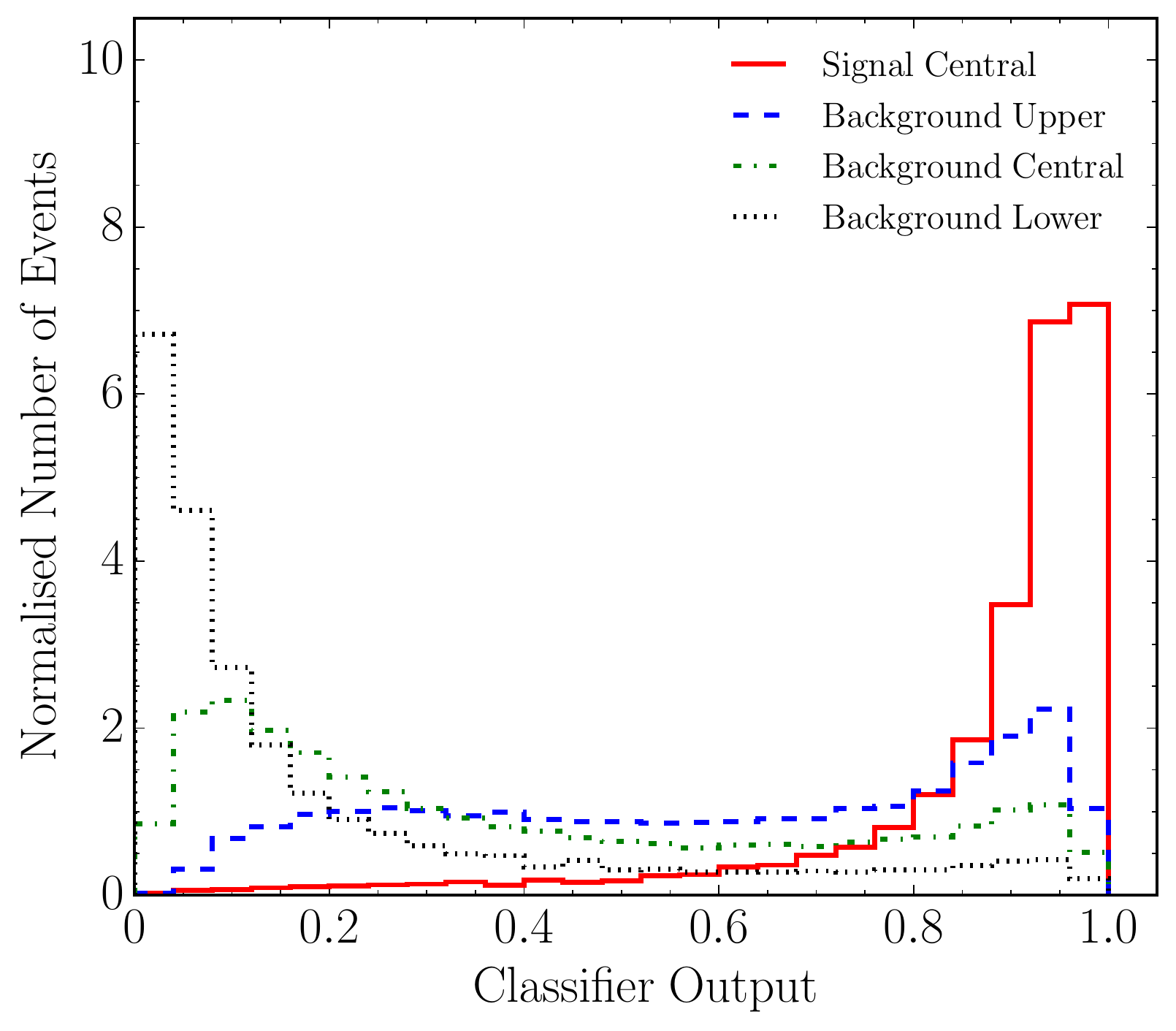}
     \caption{}
   \end{subfigure}
   \begin{subfigure}{0.49\linewidth} \centering
     \includegraphics[width=\textwidth]{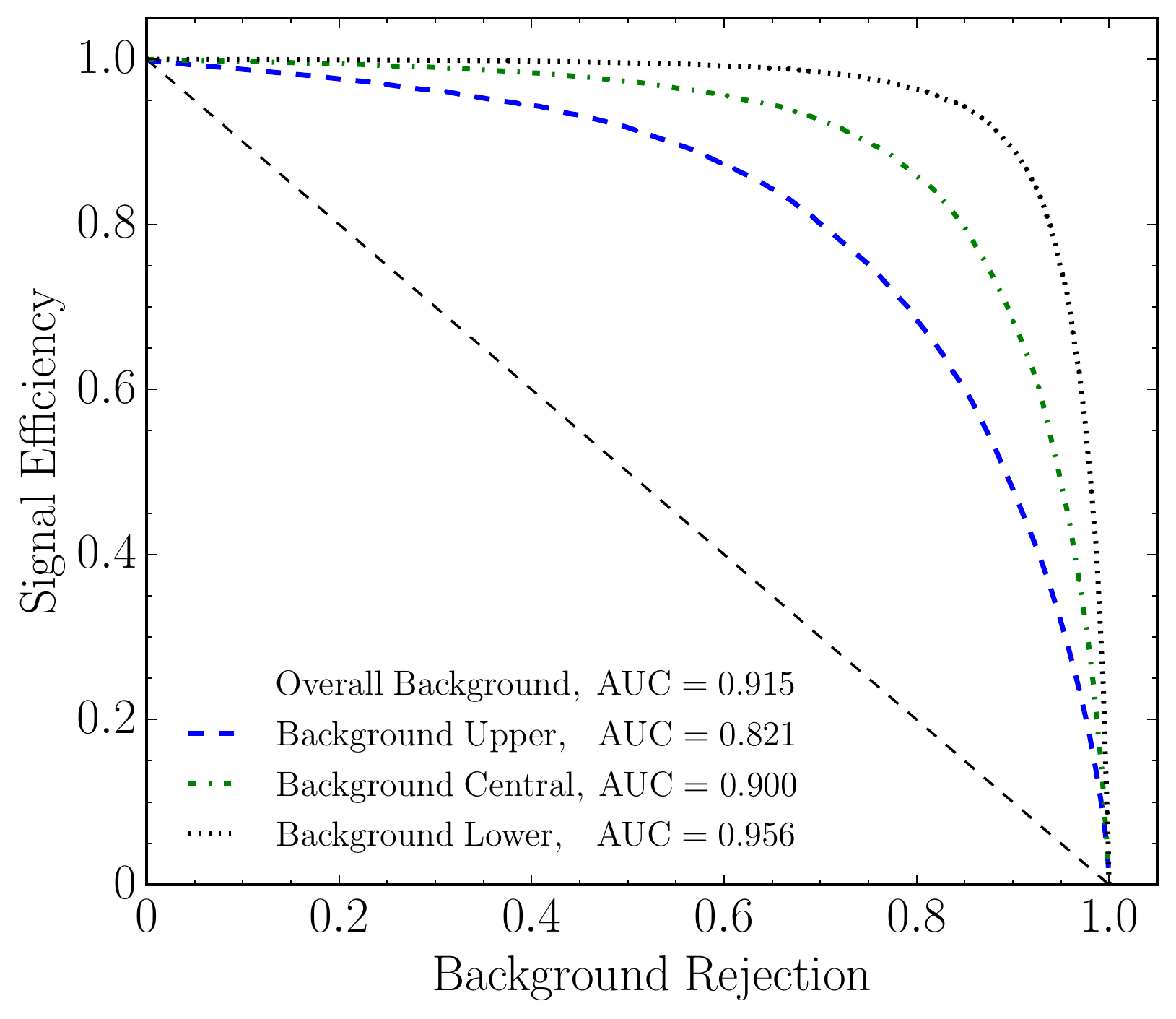}
     \caption{}
   \end{subfigure}
\caption{Supervised neural network classifier output (a) and ROC curves (b) for a classifier trained to classify signal and background events. The three background distributions result from the three different directions of smearing.} \label{fig:cls}
\end{figure}

The network consists of two hidden layers each with 20 nodes, with ReLu activations, and a final layer with a single sigmoid output. We use a binary cross entropy loss function since there are two possible classes. A class weighting in the loss function is used to account for the higher frequency of background events in the training data, i.e.~the loss of the signal events are weighted higher. The network is trained using the Adam optimiser \cite{Kingma:2014} with a learning rate of 0.01 and a batch size of 500 for 500 epochs. The network is implemented in \textsc{Keras} \cite{keras} with a \textsc{TensorFlow} \cite{tensorflow} backend, and we use these throughout the rest of this paper.  The results are shown by the distributions of the classifier outputs and the corresponding receiver operating characteristic (ROC) curves in Fig.~\ref{fig:cls}. These are obtained by testing the network on each of the three background sets separately, and performing a classification against the central signal sample for each one. Also shown are the area-under-curve (AUC) scores for each curve as well as the score for all the background test samples combined. The network performance is strongly dependent on the direction that the sample has been smeared in. This can be understood from the observables in Fig.~\ref{fig:observables} where there is a larger overlap between the signal distribution and the background which has been smeared upwards. 

We now extend this classifier with an adversarial network which is designed to discriminate the smearing class that the background sample came from, based upon the output of the classifier. The aim for such an extension to the classifier is to attempt to remove such a large dependence of its performance on the smearing of the background \cite{Louppe:2016ylz,Englert:2018cfo}. The adversary and classifier are forced to take part in a zero-sum game---the classifier must learn to make its prediction without using any information derived from the smearing, in order to make it as hard as possible for the adversary to be able to discriminate the background samples. This is achieved by the two networks having opposite optimisation objectives, so that the classifier is penalised when the adversary performs better.

The adversarial network consists of two hidden layers with 20 nodes and ReLu activation functions, and takes as an input the output of the classifier. The output of the adversary has three nodes (one for each smearing class) with a softmax activation function and a categorical cross entropy loss. The network is then trained as follows:

\begin{enumerate}
\item The classifier is trained for three epochs using the Adam optimiser with a learning rate of 0.01 and a batch size of 500. A class weighting is applied to account for the higher frequency of background events in the training data.
\item The adversary is trained on background events for three epochs using mini-batch gradient descent with a learning rate of 0.01 and a batch size of 500.
\item The classifier is trained for one epoch with mini-batch gradient descent with a batch size of 500 and with a total loss function,
\begin{equation}
\label{eq:combined_loss_class}
\mathcal{L}_{\mathrm{tot}} = \mathcal{L}_{\mathrm{class}} - \alpha \mathcal{L}_{\mathrm{adv}}~.
\end{equation}
Furthermore, two class weightings are applied: one to account for the higher frequency of background events that the classifier is trained on, and one to account for the fact that the signal events are unsmeared, resulting in a higher frequency of unsmeared events that the adversary is trained on.
\item The adversary is trained on background events for one epoch using mini-batch gradient descent with a batch size of 500.
\item Steps 3 and 4 are repeated until they have been performed a total of 1000  times, with the learning rate decaying every 100 epochs to a factor of 0.75 of its previous value, starting from an initial value of 0.01.
\end{enumerate}

\begin{figure}[!t]
\centering
   \begin{subfigure}{0.475\linewidth} \centering
     \includegraphics[width=\textwidth]{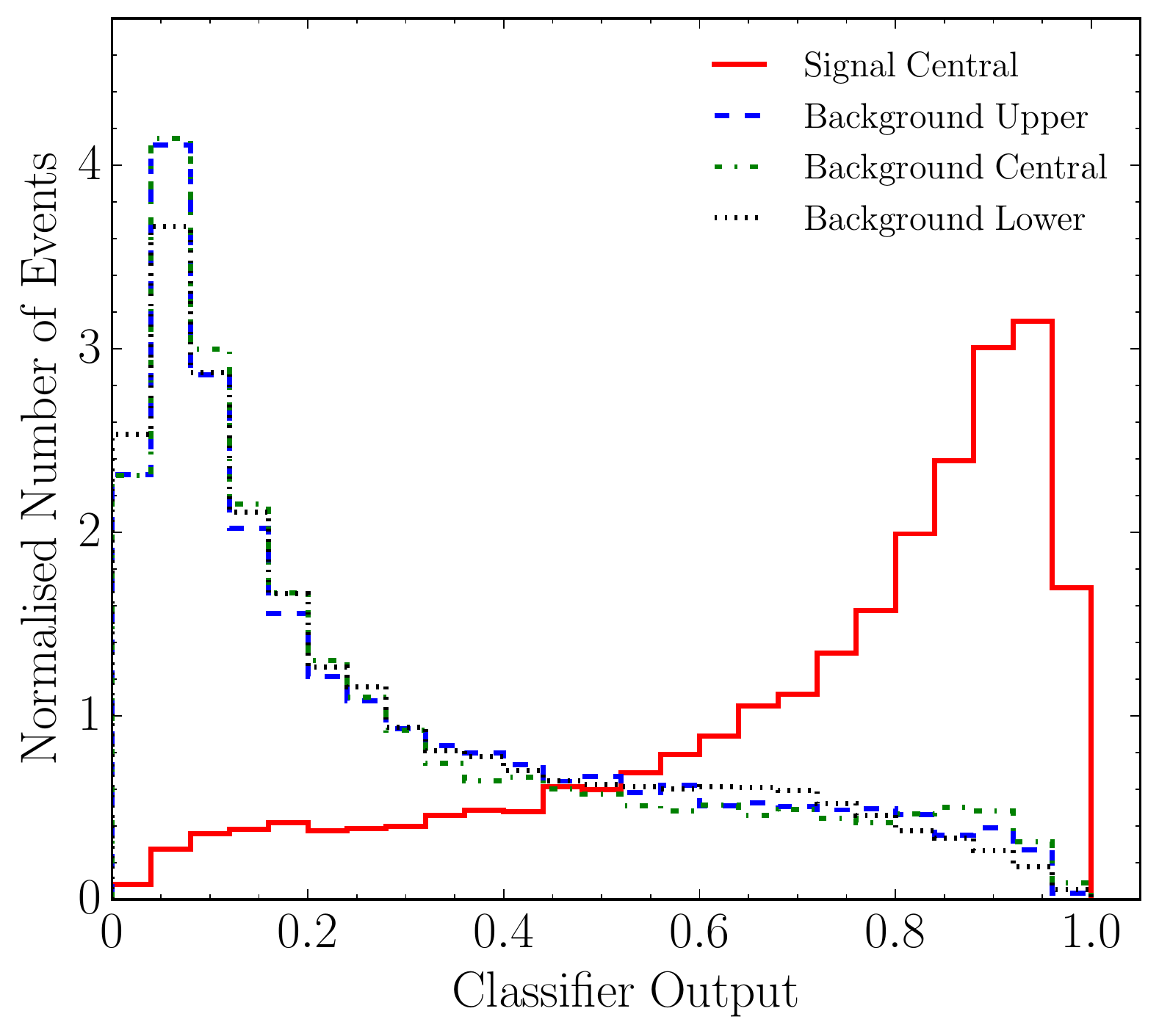}
     \caption{}
   \end{subfigure}
   \begin{subfigure}{0.49\linewidth} \centering
     \includegraphics[width=\textwidth]{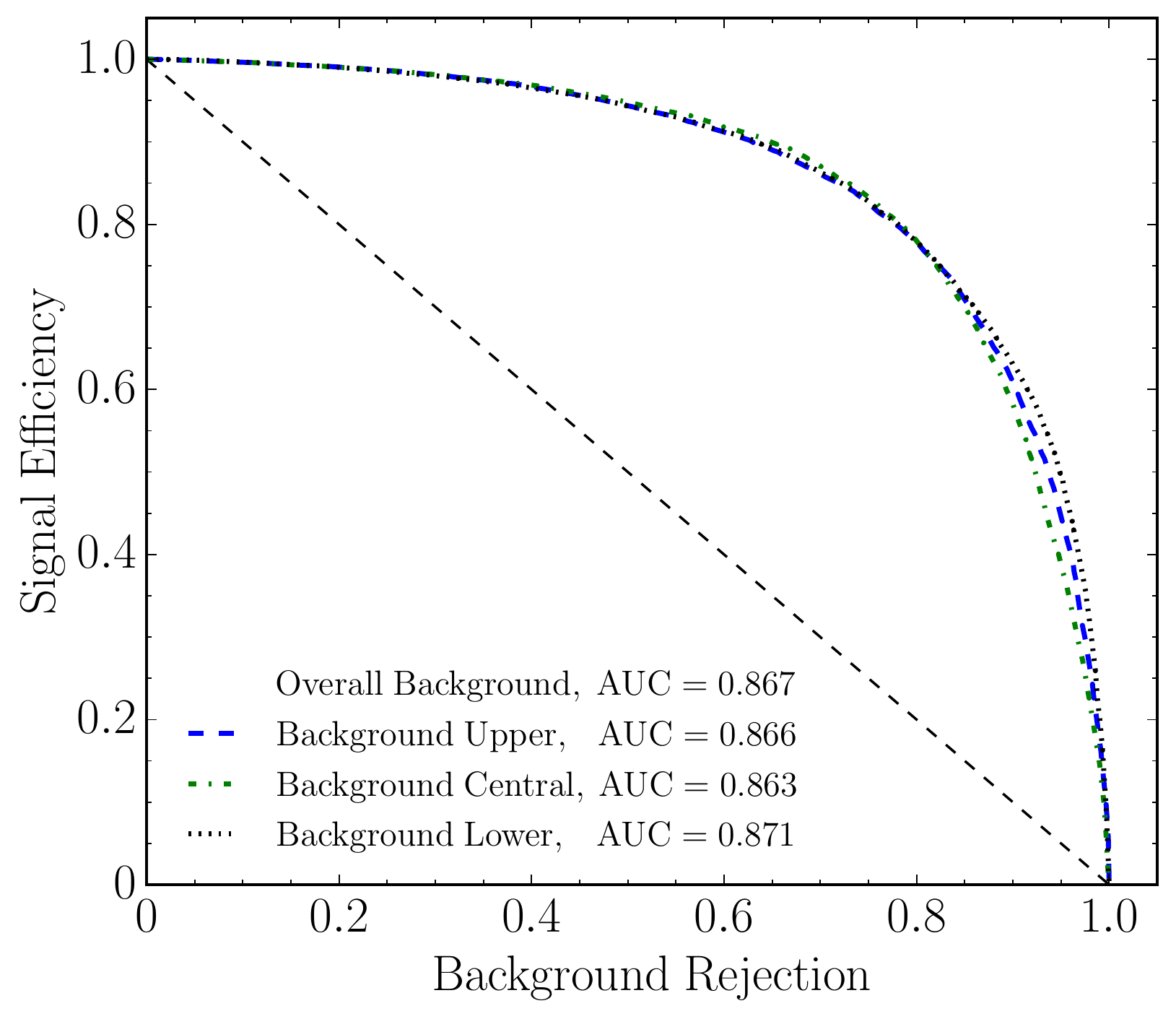}
     \caption{}
   \end{subfigure}
\caption{Supervised neural network classifier output (a) and ROC curves (b) for an adversarial classifier trained to classify signal and background events. The three background distributions result from the three different directions of smearing.} \label{fig:advcls}
\end{figure}

The weight factor $\alpha$ in Eq.~\eqref{eq:combined_loss_class} determines the relative importance of the two optimisation objectives. If it is set to zero, then the adversary has no effect on the training of the classifier. If it is too large, however, the performance of the classifier is severely affected. We find a value of 100 works well for our setup. There is another approach to training the adversarial network, where one updates the weights of both networks simultaneously. However, we find the approach of alternating the training---where the classifier is trained with the adversary weights frozen, and vice versa---to be more stable.

In Fig.~\ref{fig:advcls}, we show the performance of the adversarial classifier through the classifier output and ROC curves. The adversary has clearly had the effect of shaping the classifier outputs such that their dependence on the background smearing has been almost entirely removed. Thus, the ROC curves and AUC scores become very close together since the classification performance is now barely affected by the smearing.

\section{Extension to unsupervised autoencoder}
\label{sec:unsupervised}
\subsection{Adversarial autoencoder}

\begin{figure}[!t]
\centering
   \begin{subfigure}{0.49\linewidth} \centering
     \includegraphics[width=\textwidth]{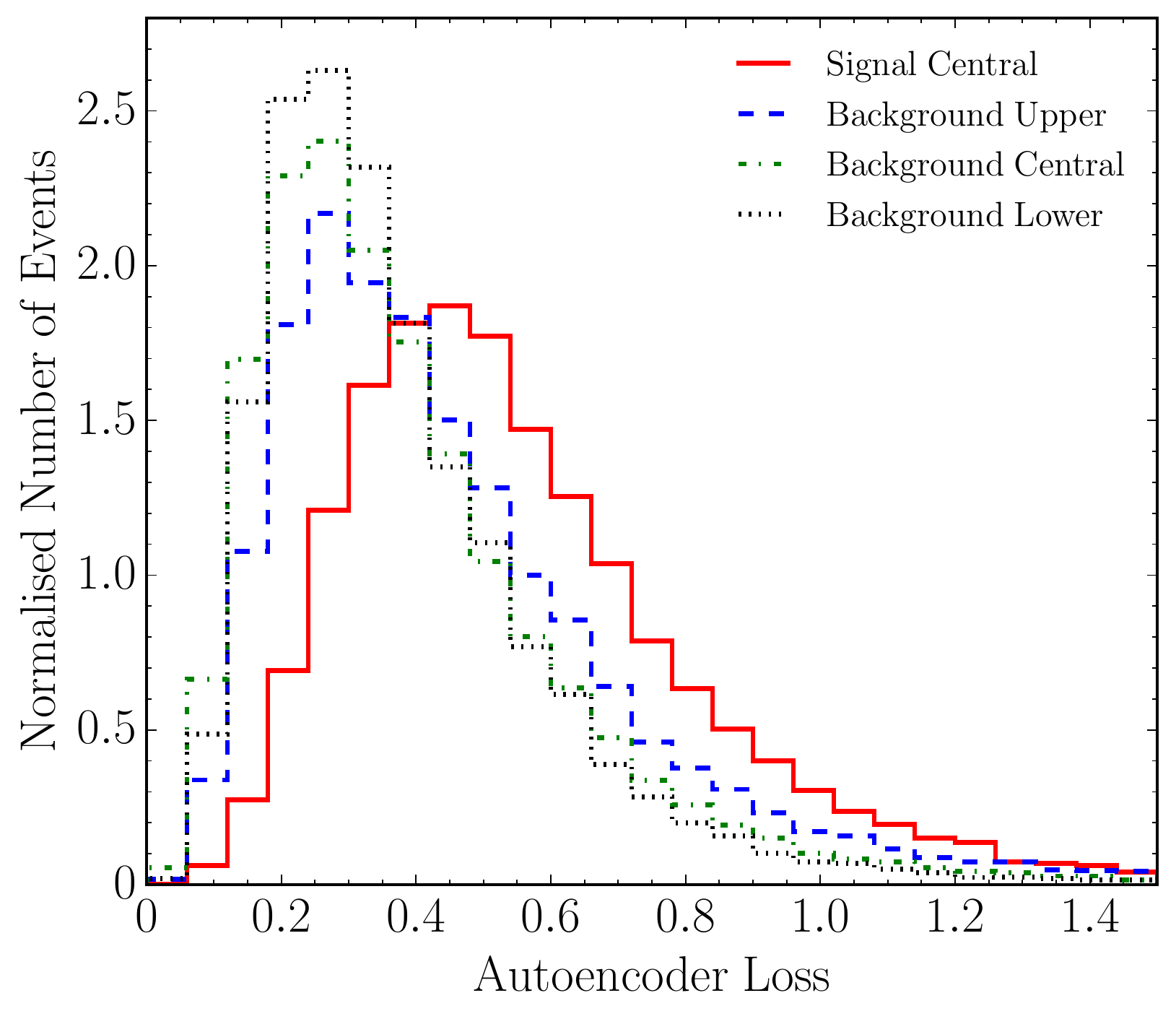}
     \caption{}
     \label{fig:AEresponse}
   \end{subfigure}
   \begin{subfigure}{0.49\linewidth} \centering
     \includegraphics[width=\textwidth]{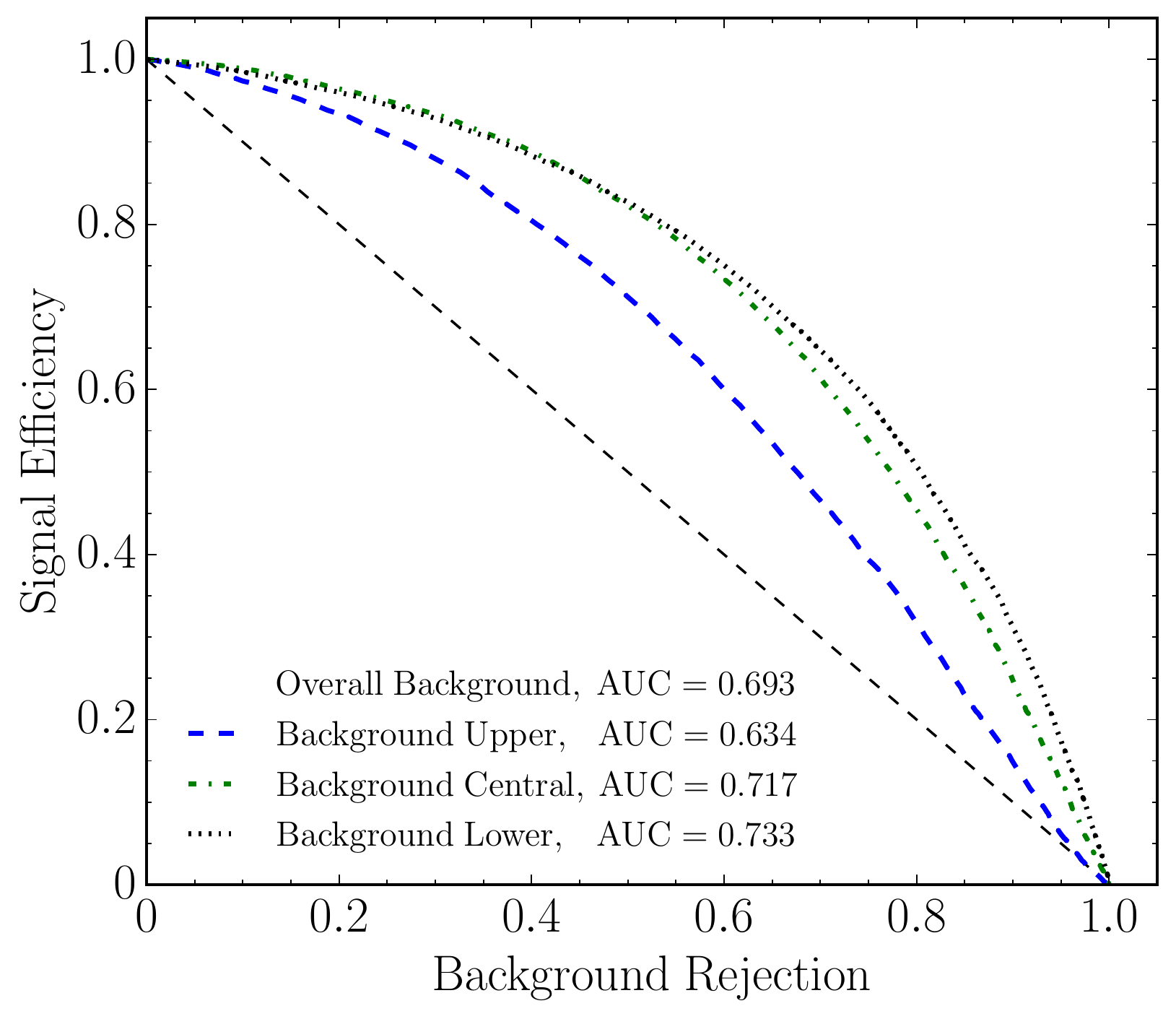}
     \caption{}
     \label{fig:AEROC}
   \end{subfigure}
\caption{Autoencoder loss (a) and ROC curves (b) for an autoencoder trained only on background events. The three background distributions result from the three different directions of smearing.} \label{fig:AE}
\end{figure}

As described earlier, autoencoders are an unsupervised learning algorithm which can be used as anomaly detectors to search for new physics since they only need to be trained on the background. 

To this aim, we consider an autoencoder constructed from three hidden layers with 10, 3 and 10 nodes respectively, each with sigmoid activation functions. After the hidden layers, there is a linear output layer with the same dimension as the number of inputs, which correspond to the 21 observables. The loss is the mean squared error between the inputs and outputs---namely, the autoencoder has the goal of reconstructing the inputs as well as possible, having encoded the information into the latent-compressed layer. We train the autoencoder on the three background samples using the Adam optimiser with a learning rate of 0.01 for 500 epochs, and the results are shown in Fig.~\ref{fig:AE}. Since the autoencoder is trained only on the background events, it learns how to reconstruct background events better than the signal events, and so the distribution of the losses for the signal events in Fig.~\ref{fig:AEresponse} is at higher values. The ROC curves in Fig.~\ref{fig:AEROC} are obtained by performing a cut on the loss function and labelling all events above the cut as signal events, and all events below the cut as background events, and then moving this threshold across all values. This is similar to how the ROC curves are calculated from the output of the classifier, where the threshold is varied between 0 and 1 instead.

As we saw for the classifier, the smearing of the background has an effect on how well the autoencoder can be used to classify events, with the events which have been smeared upwards being mislabelled as signal events more often. It is important to note that the overall classification performance of the autoencoder is much worse than for the dedicated supervised classifier in Sec.~\ref{sec:supervised}. However, this is not surprising---the autoencoder is only ever trained on background events, and only sees the signal events during testing. Thus, for the separation between signal and background it learns the intricate kinematic features of the background only. Furthermore, the optimisation objective of the classifier is for it to achieve a strong classification performance, which is not the case for the autoencoder.

\begin{figure}[!t]
\centering
     \includegraphics[width=\textwidth]{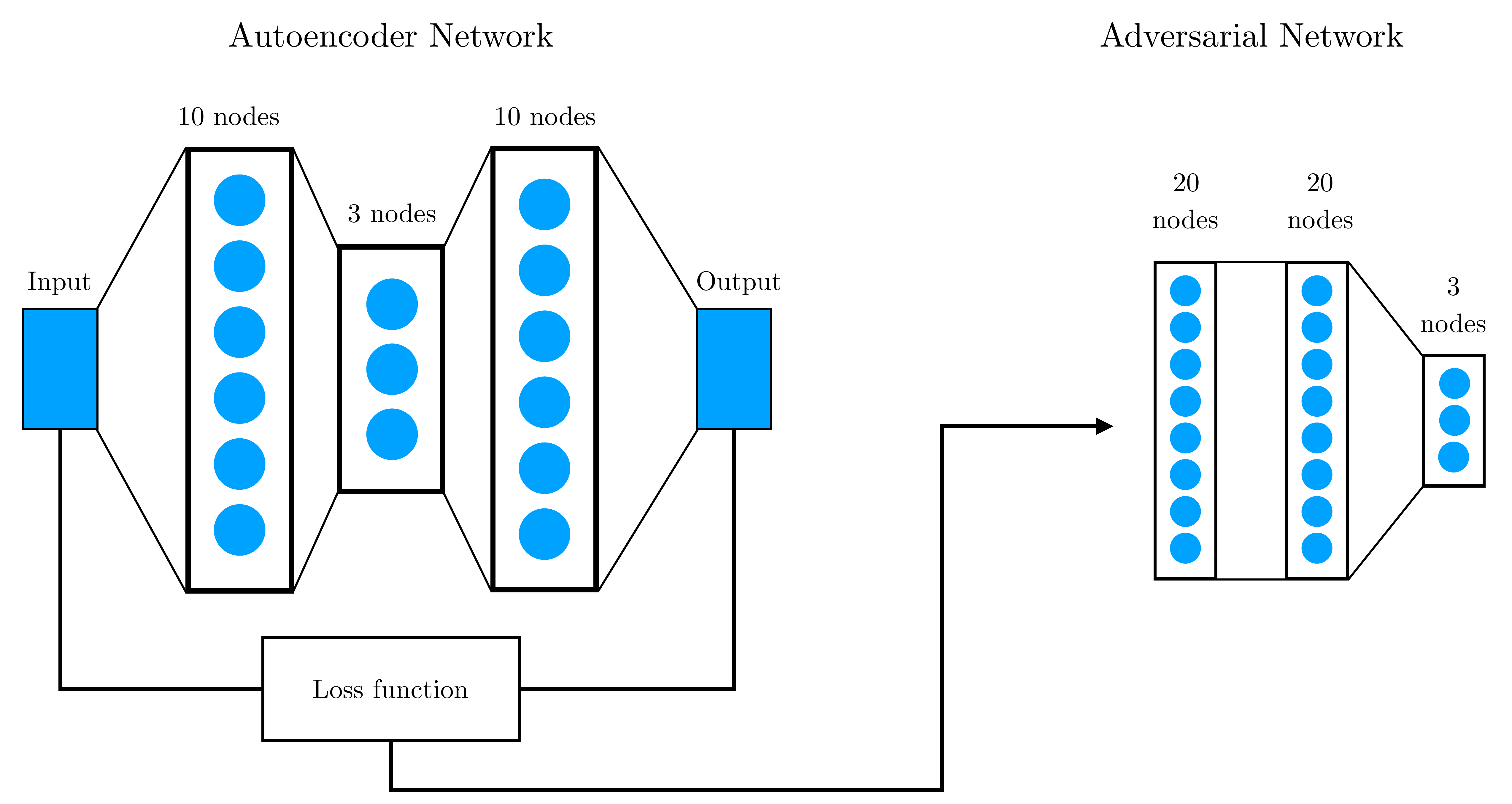}
\caption{Architecture of the adversarial autoencoder. The loss function of the autoencoder is used as an input to the adversary for it to discriminate the smeared background samples.} \label{fig:advAE}
\end{figure}

We now combine the autoencoder with an adversarial network to improve the reliability and robustness of this unsupervised-learning approach. To achieve the aim of the autoencoder being able to make its predictions independent of the smearing of the background, we use the autoencoder loss as an input to the adversary. Since a threshold on the autoencoder loss is used to perform the classification between signal and background, it is completely analogous to the output of the dedicated classifier used above, on which a cut is placed to classify the events. This input is then followed by two hidden layers each with 20 nodes and ReLu activation functions, with three softmax output nodes and a categorical cross entropy loss. This architecture is illustrated by the diagram in Fig.~\ref{fig:advAE}. The training proceeds similarly to the adversarial classifier, but with only background events in the training sample:

\begin{enumerate}
\item The autoencoder is trained for three epochs using the Adam optimiser with a learning rate of 0.01 and a batch size of 500.
\item The adversary is trained for three epochs using mini-batch gradient descent with a learning rate of 0.01 and a batch size of 500.
\item The autoencoder is trained for one epoch with mini-batch gradient descent with a batch size of 500 and with a total loss function,
\begin{equation}
\label{eq:combined_loss_AE}
\mathcal{L}_{\mathrm{tot}} = \mathcal{L}_{\mathrm{auto}} - \alpha \mathcal{L}_{\mathrm{adv}}~.
\end{equation}
\item The adversary is trained for one epoch using mini-batch gradient descent with a batch size of 500.
\item Steps 3 and 4 are repeated until they have been performed a total of 1500 times, with the learning rate decaying every 100 epochs to a factor of 0.75 of its previous value, starting from an initial value of 0.01.
\end{enumerate}

\begin{figure}[!t]
\centering
   \begin{subfigure}{0.49\linewidth} \centering
     \includegraphics[width=\textwidth]{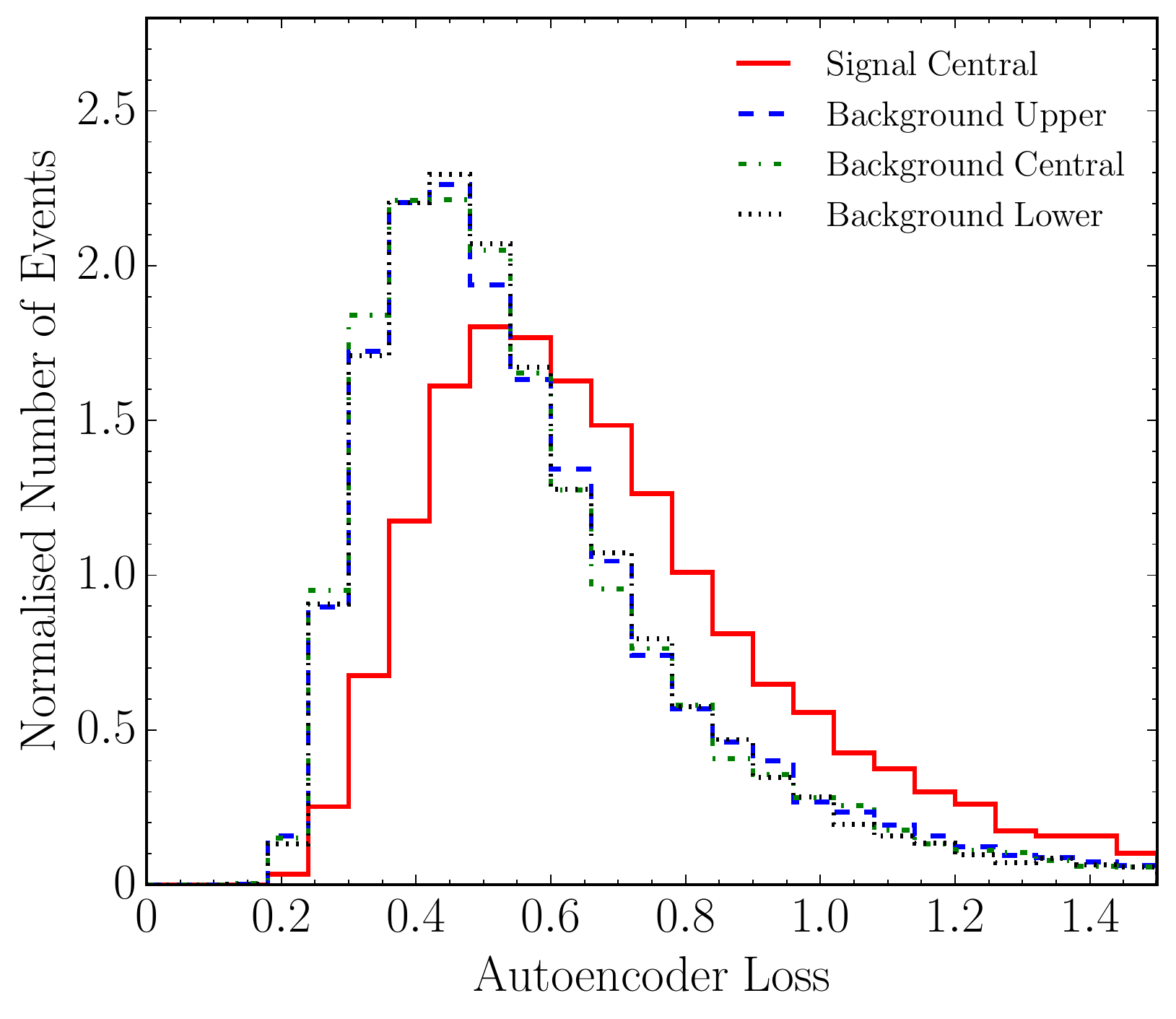}
     \caption{}
     \label{fig:AEadvresponse}
   \end{subfigure}
   \begin{subfigure}{0.49\linewidth} \centering
     \includegraphics[width=\textwidth]{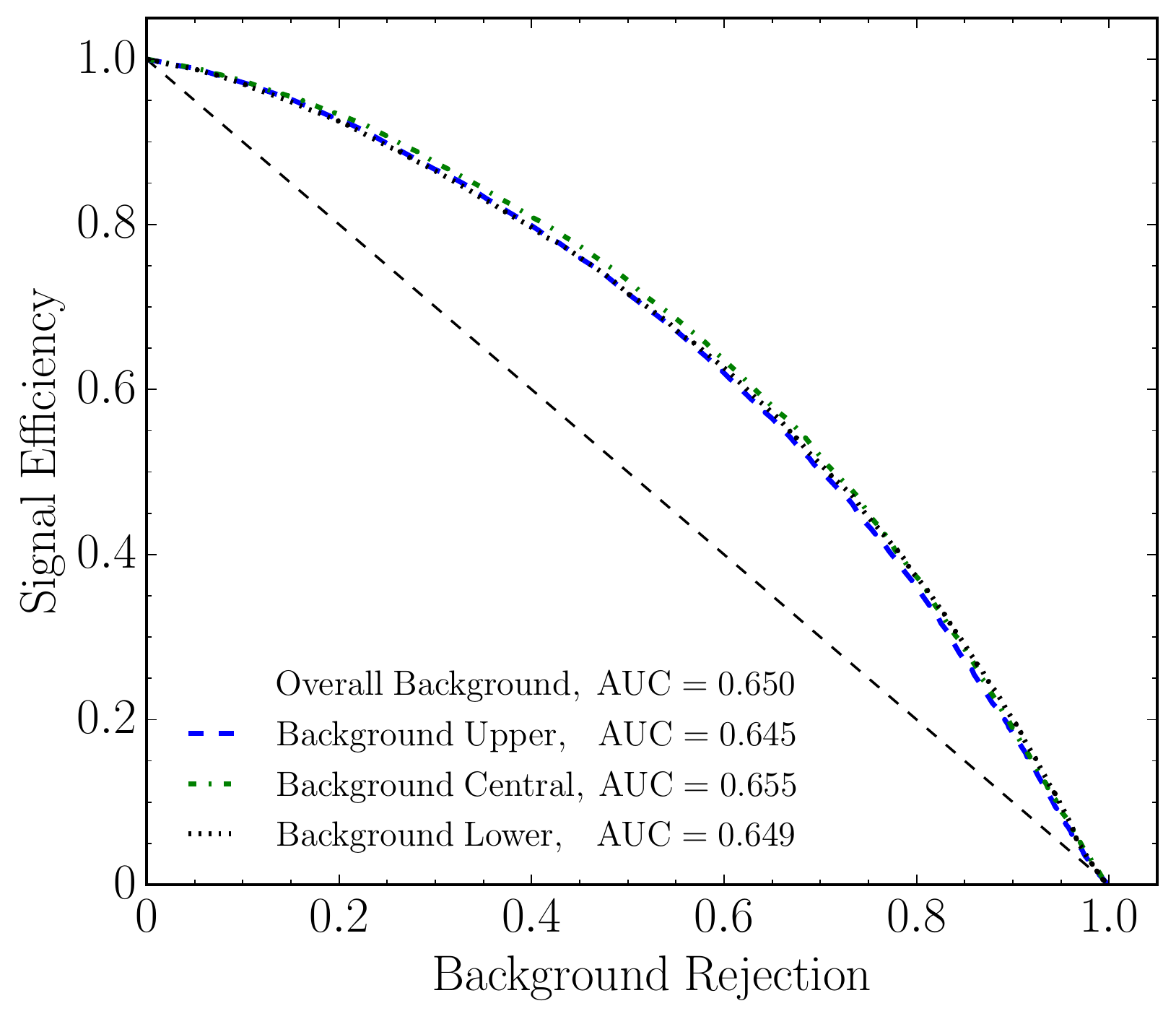}
     \caption{}
     \label{fig:AEadvROC}
   \end{subfigure}
\caption{Autoencoder loss (a) and ROC curves (b) for an adversarial autoencoder trained only on background events. The three background distributions result from the three different directions of smearing.} \label{fig:AEadv}
\end{figure}

We find this procedure to provide stable and numerically reliable results. Again, the relative weighting between the autoencoder and the adversary is set to $\alpha=100$. The performance of the adversarially-trained autoencoder is shown in Fig.~\ref{fig:AEadv}. The background distributions shown in Fig.~\ref{fig:AEadvresponse} have been shaped such that they are independent of the direction of smearing, which results in the ROC curves in Fig.~\ref{fig:AEadvROC} becoming almost identical. This shows that the method has become independent of uncertainties inherent to the reconstruction of the final-state objects of LHC events.

In addition, we note that our setup also has the ability to interpolate to smaller amounts of smearing---although we have trained using background data which has been systematically smeared by a very large amount, we find that if it is tested on samples which have been smeared by a much smaller amount (without the increase by a factor of three), then the output of the adversarially-trained autoencoder (and also for the classifier in the previous section) is still insensitive to the smearing. Furthermore, we find that the AUC score increases when it is tested on a smaller amount of smearing, and is similar to the result of having both trained and tested it on this smaller amount. Therefore, the fact that the adversary is trained on a larger amount of smearing than is realistic does not adversely affect its performance.

We will now briefly recap what we have achieved by combining an autoencoder with an adversarial neural network. We started with three sets of background events---one which had been smeared upwards, one which had been smeared downwards, and one which had not been smeared at all. This smearing corresponded to the extremities of a Gaussian envelope, and was applied to jets, leptons and the missing energy in each event accordingly. Furthermore, we also had a set of signal events which had not been smeared. The smearing had the effect of shifting the kinematic features of the background such that the events which had been smeared upwards looked more like signal events, and the ones which had been smeared downwards looked less like signal events. This can be seen from the distributions in Fig.~\ref{fig:observables}.

We then trained an autoencoder on all the background events for the purpose of using it to detect signal events, which have a higher expected reconstruction loss. In Fig.~\ref{fig:AEROC}, the ROC curves are the result of testing the classification performance of the autoencoder for the signal separately against each background, and as expected, the autoencoder had a harder time discriminating the signal events against background events which had been smeared upwards. We then combined this with an adversarial neural network, which had the objective of recognising which direction each background sample had been smeared in based upon the loss of the autoencoder. The autoencoder and adversary were trained using a combined loss function, which penalised the autoencoder for outputting reconstruction losses from which the adversary could discriminate the samples. The result of this is that the autoencoder has learnt to reconstruct events without using any information derived from the smearing, which can be seen from the fact that the ROC curves in Fig.~\ref{fig:AEadvROC} have converged.

\subsection{Corrupted autoencoder and application to other new physics models}

\begin{figure}[!t]
\centering
     \includegraphics[scale=0.5]{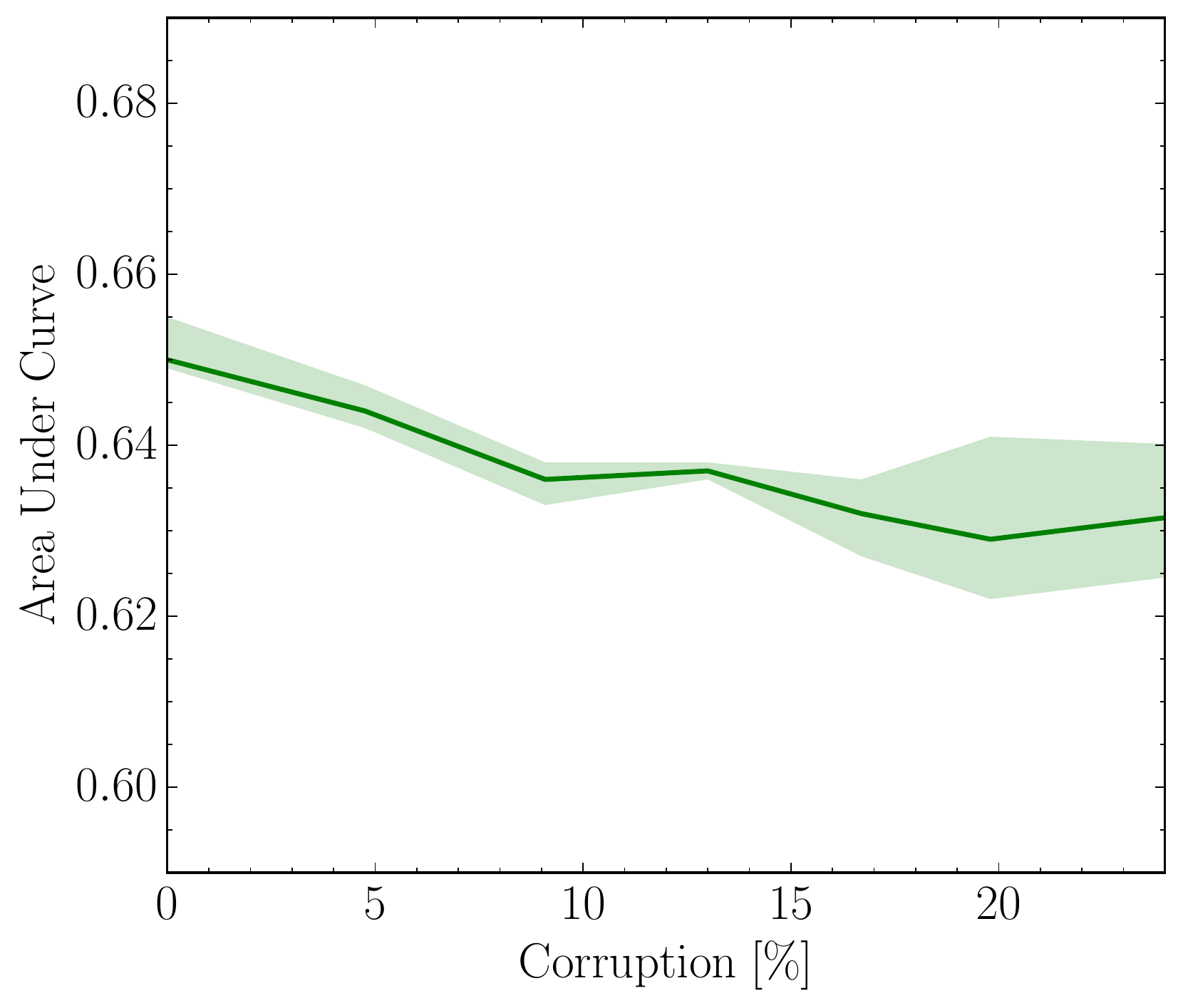}
\caption{Effect of contaminating the training sample with an increasing fraction of signal events. The central line shows the overall AUC score, and the band represents the difference between the upper and lower AUC scores.} \label{fig:corruption}
\end{figure}

Thus far, the analysis has been carried out on training sets consisting of pure background events. Realistically, data may not actually look like this since if new physics exists, then it would also form part of that same data. To begin accounting for this it is possible to inject into the three background sets appropriately smeared signal events. By training on these newly contaminated sets we can investigate how sensitive the performance of the adversarial autoencoder is to an increase in signal corruption in the training set. In Fig.~\ref{fig:corruption}, we show these results. The band represents the difference between the upper and lower AUC scores, which shows how well the adversary desensitises the autoencoder from the smearing, and the central line is the overall AUC score. All model hyperparameters are left unchanged during the training, with only the relative fraction of corruption changing, defined as a percentage of the total training set. From the plot it is clear that injecting signal events during training has little effect on the overall performance until the fraction of corruption becomes unrealistically large, showing the potential applicability of the method to real data.

Since the performance is not drastically affected by a corruption of the training data, we can proceed with a training sample consisting purely of background events. One of the advantages of the autoencoder only needing to be trained on background events is that it can then be tested for signal events arising from any model. Here, we test our adversarially-trained autoencoder on a variety of different new physics models. We aim to quantify the effect of the resonance's quantum numbers, i.e.~spin, colour and coupling strengths, on the performance of the autoencoder. The models used are: 
\begin{itemize}
  \item  Two further $Z^\prime$  cases with widths of $10$~GeV and $200$~GeV. In both cases the masses are held at $2$~TeV.
  \item A scalar colour-octet \cite{Frederix:2007gi}, with a mass of $2$~TeV and the scalar and axial parameters fixed to ensure the width is $\sim89.6$~GeV.
  \item A scalar colour-singlet with a mass of $2$~TeV and a width of $89.6$~GeV.
\end{itemize}

Table~\ref{tab:bsmResultsAE} shows the results of testing the adversarially-trained autoencoder on the new signals. In each case the adversary is able to perform well, with the difference between the upper and lower AUC scores showing that the new signals do not hinder the ability of adversary to desensitise the autoencoder to the smearing. This behaviour is of course expected, since the same background samples are used to test against each new signal. We also show estimates of the potential limits on the cross sections that can be obtained using the classification performance of the autoencoder. These are calculated by finding the points on the ROC curves that maximise $S/\sqrt{B}$, then comparing them to the background cross section and assuming an integrated luminosity of 100~fb$^{-1}$. We then require that $S/\sqrt{B}>2$ to set a 95\% confidence limit. The limits we find are insensitive to the nature of the resonance i.e.~with respect to their quantum numbers, and they are comparable to the limits found by ATLAS in Ref.~\cite{Aaboud:2018mjh}.\footnote{However, note that we show the new physics cross section after event selection and reconstruction cuts, while ATLAS shows the inclusive cross section for a specific $Z^\prime$ model. Furthermore, our analysis was performed at 14 TeV, while the limits from ATLAS have been obtained at a centre-of-mass energy of 13 TeV.}

\begin{table}[]
\centering
\begin{adjustbox}{width=\columnwidth}
\begin{tabular}{c|c|c|c}
\textbf{Signal} & \textbf{Overall AUC}& \textbf{Upper-Lower Difference} & \textbf{Cross Section Limit [pb]} \\ \hline
$Z^\prime_{w=10~\mathrm{GeV}}$     & 0.662                          & 0.009        & 0.0101                   \\ \hline
$Z^\prime_{w=89.6~\mathrm{GeV}}$    & 0.656                          & 0.009         & 0.0098                  \\ \hline
$Z^\prime_{w=200~\mathrm{GeV}}$   & 0.650                           & 0.009           & 0.0105               \\ \hline
Scalar           & 0.654                         & 0.010                & 0.0104           \\ \hline
Octet           & 0.659                            & 0.010             & 0.0102             
\end{tabular}
\end{adjustbox}
\caption{The overall AUC score, difference between the largest and smallest AUC scores and the cross section limits found from using the adversarial autoencoder trained only on background events and tested on the original $Z^\prime$ case and four other signals. }
\label{tab:bsmResultsAE}
\end{table}

\section{Conclusions}
\label{sec:conclusions}

The ideal scenario for the usage of machine learning methods is when they can be applied directly on experimental data, without the requirement to train them on pseudo-data or without theoretically calculated inputs, e.g.~as in the Matrix Element Method. In such circumstances neither theoretical uncertainties that challenge the robustness of the method, nor a theoretical bias regarding the features of the signal are introduced. Thus, the experimental data alone would be sufficient to identify anomalous events, which could be isolated and studied further to discover new physics. Such identification of anomalous events can be realised using an autoencoder, trained on a pure background sample. However, even in this ideal scenario, residual uncertainties due to the imperfect reconstruction of final-state objects remain. 

Focusing on resonance searches in semileptonic $t\bar{t}$ final states, we quantified the performance of an adversarially-trained autoencoder. In particular, we compared the performance of an autoencoder-based unsupervised-learning approach with a supervised neural network classifier. While the supervised classifier performs significantly better than the unsupervised-learning approach, the latter still shows a strong aptitude in telling apart signal from background events. In both cases reconstruction uncertainties show however a big impact on the evaluated performance of the classifiers, thereby evidencing the need for measures to desensitise them against such uncertainties for a reliable performance evaluation. 

We proposed to combine the autoencoder with an adversarial neural network to realise a robust and reliable unsupervised anomaly detection method that can be readily applied to experimental data. The classification result is independent of the smearing of the reconstructed final-state objects over the entire range of the ROC curve and even extends to training on corrupted backgrounds, i.e.~backgrounds with a large admixture of signal events. Although we applied it to Monte-Carlo-generated pseudo-data, we envisage that the procedure could be applied analogously to experimental data by creating labelled datasets that have been systematically smeared. Thus, this setup proves to be a very robust data-driven way to search for new physics resonances, irrespective of their quantum numbers, i.e.~spin, colour or width.

\vskip 1 \baselineskip

\noindent {\it{Acknowledgements.  MS acknowledges the generous hospitality of Barbara Jaeger and her group at the University of Tuebingen, as well as support of the Humboldt Society, during the completion of parts of this work.}}

  
\bibliographystyle{JHEP}
\bibliography{references}

\end{document}